\documentclass[12pt]{article}
\usepackage{latexsym}
\usepackage{amssymb}
\usepackage{amsmath}
\usepackage{graphicx}
\newcommand{\phis}{$\langle\phi^{2}\rangle$}
\newcommand{\texp}{$\langle T_{\mu \nu}\rangle$}

\begin{document}

\title{\phis\ in the Spacetime of a Cylindrical Black Hole}
\author{\small Andrew DeBenedictis \footnote{e-mail:adebened@sfu.ca} 
\\ \it{\small Department of Physics} 
\\ \it{\small Simon Fraser University, Burnaby, B.C., Canada V5A 1S6}}
\date{March 23, 1998}
\maketitle

\begin{abstract}
The renormalised value of \phis\ is calculated for a massless, 
conformally coupled scalar field in the Hartle-Hawking vacuum state. This 
calculation is a first step towards the calculation of the gravitational 
back reaction of the field in a black cosmic string spacetime which is 
asymptotically anti-DeSitter and possesses a non constant dilaton field. 
It is found that the field is divergence free throughout the spacetime 
and attains its maximum value near the horizon. 
\end{abstract}

\section{Introduction}

\qquad A useful question to ask when one studies quantum fields in 
General Relativity is the following: Given that all matter is inherently 
quantum in nature, will quantum effects remove the singularity at the 
centre of a black hole spacetime? To answer this question using a scalar 
field and 
semi-classical perturbation theory, one must first calculate the expectation 
value \texp\ where $T_{\mu \nu}$ is the stress-energy tensor operator of the 
scalar field $\phi$. In this paper \phis\ is computed in preparation for the 
computation of \texp\ which is used as the source term for the Einstein 
field equations:

\begin{equation}
R_{\mu\nu}-\frac{1}{2}Rg_{\mu\nu}-\Lambda g_{\mu\nu}
=8\pi \langle T_{\mu\nu}\rangle . 
\label{eq:einst} \end{equation}
A review of handling quantum fields in the presence of strong 
gravitational fields can been found in articles by Wipf ~\cite{wipf} and 
DeWitt ~\cite{dewitt}  and in books by Birrel and Davies 
~\cite{biranddav} and Wald ~\cite{waldbook}.

\qquad \phis\ for massless fields has been computed for both the interior
and exterior of a Schwarzschild black hole ~\cite{ch:schw}
~\cite{cj:schw}.  These calculations have also been extended by Anderson to 
accommodate
massive fields in general spherically symmetric, asymptotically 
flat spacetimes~\cite{and:mass}. A method has also been developed by 
Anderson, Hiscock and Samuel ~\cite{ahs1} ~\cite{ahs2} to calculate    
the expectation value of the stress-energy operator, \texp\ in 
spherically symmetric, static spacetimes. They use this method to 
calculate \texp\ in Schwarzschild and Reissner-Nordstr\"{o}m geometries.
\texp\ has also previously been computed by Howard and 
Candelas ~\cite{hc:schw} in Schwarzschild spacetime. 

\qquad The Kerr spacetime has also been studied with fields propagating 
in this geometry. Frolov ~\cite{fr:kerr1} has calculated 
\phis\ for massless fields on the event horizon pole of a Kerr-Newmann 
black hole as well as deriving an approximate expression for \texp\ near the 
horizon with Thorne ~\cite{frtho:kerr3}. Massive fields in the exterior 
geometry have been studied by Frolov and Zel'nikov ~\cite{frzel:kerr4}.

\qquad Various works on back reaction effects of quantum fields on black 
hole geometries have been produced. For Schwarzschild and 
Reissner-Nordstr\"{o}m spacetimes this includes the work of Hiscock and 
Weems ~\cite{hw:back}, Bardeen and Balbinot ~\cite{bd:back} 
~\cite{bb:back}, and York ~\cite{yk:back} who used Page's analytic 
approximation ~\cite{page} for \texp\ in Einstein spacetimes for 
conformally invariant 
fields. More recently Hiscock and Larson ~\cite{hiscock} have extended their 
analysis to the Schwarzschild interior and calculated back reaction 
effects on curvature invariants.

\qquad Most work in this field has been done in the context of spherical 
or oblate symmetry. We wish to extend the analysis to other symmetries and 
ultimately ask if the above results are general or are specific to the 
particular symmetry chosen. For example, does the presence of the quantum 
field have the same effect on the curvature growth and on the 
anisotropy of a black hole interior for all symmetries? If curvature 
invariants are weakened for all systems studied then one can say with 
some confidence that quantum effects may remove the singularity. 

\subsection{Background geometry}

\qquad The system studied here will be that of a massless Klein-Gordon
field with conformal curvature coupling propagating in the spacetime
generated by a straight black string. This background is chosen for several
reasons. First, it possesses cylindrical (as opposed to spherical) symmetry
and secondly, in the context of cosmic strings it represents a system
which may physically exist in the universe. It could also be argued that
sufficiently close to a black string loop the spacetime will possess
this type of geometry. There has also been a revived interest in 
anti-DeSitter spacetimes in the context of conformal field theories. The 
metric chosen is that developed by Kaloper ~\cite{kaloper} and Lemos and
Zanchin ~\cite{lemos} which admits a line element of the following form:
\begin{eqnarray}
ds^2=&-\left(\alpha^2\rho^2-\frac{2(M+\Omega)}{\alpha\rho}+
\frac{4Q^2}{\alpha^2\rho^2}\right)\; dt^2
-\frac{16J}{3\alpha\rho}\left(1-\frac{2Q^2}{(M+\Omega)\alpha\rho}\right)
dt\,d\varphi \nonumber \\ &+\left[\rho^2+\frac{4(M-\Omega)}{\alpha^3\rho}
\left(1-\frac{2Q^2}{(M+\Omega)\alpha\rho}\right)\right]d\varphi^2
\nonumber \\ &+\frac{d\rho^2}{\alpha^2\rho^2-
\frac{2(3\Omega-M)}{\alpha\rho}+\frac{(3\Omega-M)4Q^{2}}
{(\Omega+M)\alpha^2\rho^2}} +\alpha^2\rho^2\; dz^2 \label{eq:genmetric}
\end{eqnarray} 
Where $M$, $Q$, and $J$ are the mass, charge, and angular
momentum per unit length of the string respectively. $\Omega$ is given by:
\begin{equation} 
\Omega=\sqrt{M^2-\frac{8J^{2}\alpha^{2}}{9}}.
\end{equation} 
The constant $\alpha$ is defined as follows:
\begin{equation} 
\alpha^2=-\frac{1}{3}\Lambda 
\end{equation} with
$\Lambda$, the cosmological constant, negative giving the spacetime its
asymptotically anti-DeSitter behaviour. The spacetime has a well defined
time-like killing vector field with respect to which modes can be defined. 

\qquad Much interesting work has been done regarding effects of quantum 
fields in the 3D BTZ ~\cite{btz} black hole 
~\cite{oritz},~\cite{steif},~\cite{shir1},~\cite{shir2},~\cite{shir3}, 
\cite{mann}. 
The system here however does not dimensionally reduce to the 3D BTZ black 
hole whose four dimensional counterpart (without charge or angular 
momentum) is given for reference as ~\cite{lemos2}
\begin{equation}
ds_{BTZ_4}^{2}=-(\alpha^2 r^2-8M)dt^2+\frac{dr^2}{(\alpha^2 
r^2-8M)}+r^2d\varphi^2+dz^2. \label{eq:btzblack}
\end{equation}
This is due to the fact that the dilaton field is non zero and non 
constant in the corresponding three dimensional action.

\qquad In the spacetime 
considered here, both charge and angular momentum are zero yielding the 
following for (\ref{eq:genmetric})
\begin{equation}
ds^{2}=-(\alpha^{2}\rho^{2}-\frac{4M}{\alpha\rho})\,dt^{2}+
\frac{d\rho^{2}}{(\alpha^{2}\rho^{2}-\frac{4M}{\alpha\rho})}+
\rho^{2}d\varphi^{2}+\alpha^{2}\rho^{2}\,dz^{2}. \label{eq:metric}
\end{equation}
It can immediately be seen from (\ref{eq:metric}) that the spacetime 
behaves as anti-DeSitter as $\rho\rightarrow\infty$ and therefore the 
spacetime is not globally hyperbolic. This fact requires boundary 
conditions to be imposed to control the flow of information in and out of 
the time-like infinity. It has been shown ~\cite{isham} that three 
natural boundary conditions arise in ADS$_{4}$ spacetime and the condition 
used here is that of a "transparent" boundary condition.

\section{Green Function and \phis} 

\qquad In this section a Euclidean space approach is used to calculate the 
Green function. The calculation is similar to that done by 
Anderson~\cite{and:mass} who calculated \phis\ in a Reissner-Nordstr\"{o}m 
spacetime however, the method is extended here to a system which is neither 
spherically symmetric nor asymptotically flat. 

\qquad The Euclidean space method amounts to making the transformation 
$t\rightarrow\imath\tau$ in metric (\ref{eq:metric}). The quantity \phis\ 
is then defined as
\begin{equation}
\langle\phi^{2}\rangle=\lim_{x\rightarrow x'}G_{E}(x,x'),
\end{equation}
where $G_{E}(x,x')$ is the Euclidean space Green function satisfying the 
equation
\begin{equation}
\left[\Box^{2}-m^{2}-\xi R(x)\right]G_{E}(x,x')=
\frac{-\delta^{4}(x,x')}{\sqrt{g(x)}}. \label{eq:kg}
\end{equation}
The mass term in the Klein-Gordon operator will be set to zero in order 
to comply with the transparent boundary conditions mentioned earlier.

\qquad The presence of three killing fields allows the right-hand-side of 
the above equation to be expanded in terms of cylindrical functions as 
follows:
\begin{equation}
\frac{\delta^{4}(x,x')}{\sqrt{g(x)}}=
\delta(\tau-\tau')\:\frac{\delta(\rho-\rho')}{\alpha\rho^{2}}
\:\delta(\varphi-\varphi')\:\delta(z-z'),
\end{equation}
with
\begin{eqnarray}
\delta(\tau-\tau')&=&T\sum_{n=-\infty}^{\infty}e^{\imath n 2\pi 
T(\tau-\tau')} \nonumber \\
\delta(\varphi-\varphi')&=&\frac{1}{2\pi}\sum_{l=-\infty}^{\infty} 
e^{\imath l (\varphi-\varphi')} \\
\delta(z-z')&=&\frac{1}{2\pi}\int_{-\infty}^{\infty}dk\:e^{\imath 
k(z-z')}. \nonumber
\end{eqnarray}
Where $T$ is the temperature of the field. The Green function can similarly 
be expanded giving
\begin{equation}
G_{E}(x,x')=\frac{T}{4\pi^{2}}\sum_{n=-\infty}^{\infty}e^{\imath n 2\pi
T(\tau-\tau')}\sum_{l=-\infty}^{\infty}
e^{\imath l (\varphi-\varphi')}\int_{-\infty}^{\infty}dk\: e^{\imath
k(z-z')}\chi(\rho,\rho').
\end{equation}
The function $\chi(\rho,\rho')$ must satisfy the following equation
\begin{eqnarray}
\frac{d^{2}\chi(\rho,\rho')}{d\rho^{2}} 
\left(\alpha^{2}\rho^{2}-\frac{4M}{\alpha\rho}\right)+
\frac{d\chi(\rho,\rho')}{d\rho}
\:4\left(\alpha^{2}\rho-\frac{M}{\alpha\rho^{2}}\right) \nonumber \\
-\chi(\rho,\rho')\left[\frac{l^2}{\rho^{2}}+\frac{k^2}{\alpha^2\rho^2}
+\frac{n^{2}4\pi^{2}T^{2}}
{\left(\alpha^{2}\rho^{2}-\frac{4M}{\alpha\rho}\right)} 
+\xi R\right]=\frac{-\delta(\rho,\rho')}{\rho^2\alpha}. \label{eq:evolve}
\end{eqnarray}
A solution is assumed of the form
\begin{equation}
\chi(\rho,\rho')=C\Psi_{1}(\rho_{<})\Psi_{2}(\rho_{>}),
\end{equation}
where $\rho_{<}$ and $\rho_{>}$ represent the lesser and greater of $\rho$ and 
$\rho'$ respectively.
Integrating across the $\delta$ function gives the Wronskian normalisation 
condition:
\begin{equation}
C\left[\Psi_{1}(\rho_{<})\frac{\partial\Psi_{2}(\rho_{>})}{\partial\rho}-
\Psi_{2}(\rho_{>})\frac{\partial\Psi_{1}(\rho_{<})}{\partial\rho}\right]
=\frac{-1}{\alpha\rho^{2}(\alpha^2\rho^2-\frac{4M}{\alpha^2\rho^2})}.
\label{eq:wronsk}
\end{equation}

\qquad The asymptotic behaviour of the solution can be found by studying
the solutions to (\ref{eq:evolve}) in the appropriate regimes. For
$\rho=\infty$ the solution has the form:  
\begin{eqnarray}
\Psi_{1}&\sim&\rho^{-\frac{3}{2}}\rho^{\frac{\sqrt{9+\frac{4\xi
R}{\alpha^2}}}{2}} \nonumber \\
\Psi_{2}&\sim&\rho^{-\frac{3}{2}}\rho^{-\frac{\sqrt{9+\frac{4\xi
R}{\alpha^2}}}{2}}. \label{eq:infbehav} 
\end{eqnarray} 
Whereas near the horizon $\left(\rho=\frac{(4M)^{1/3}}{\alpha}\right)$ 
solutions are found to behave as:  
\begin{eqnarray} 
\Psi_{1}&\sim&\: e^{\kappa
n\int{\frac{\alpha\rho\,d\rho}{\alpha^3\rho^3-4M}}} \nonumber \\ 
\Psi_{2}&\sim&\:
e^{-\kappa n\int{\frac{\alpha\rho\,d\rho}{\alpha^3\rho^3-4M}}}. 
\label{eq:horizbehav}
\end{eqnarray} 
Where $\kappa=2\pi T$. It can easily be seen that
$\Psi_{1}$ diverges at infinity and $\Psi_{2}$ is divergent at the
horizon. The general solution with the correct asymptotic behaviour can be
found by ansatz. Such a solution takes the form 
\begin{eqnarray}
\Psi_{1}&=&\frac{1}{\sqrt{\rho^3\alpha^3\,X}}\,
e^{\int{\rho^2\alpha^3\,X\,d\rho\sqrt{\frac{g_{11}}{g_{00}g_{33}}}}} 
\nonumber \\ 
\Psi_{2}&=&\frac{1}{\sqrt{\rho^3\alpha^3\,X}}\,
e^{-\int{\rho^2\alpha^3\,X\,d\rho\sqrt{\frac{g_{11}}{g_{00}g_{33}}}}}.
\label{eq:ansatz} 
\end{eqnarray} 
$X$ is a function of $\rho$ which evolves according to the equation:
\begin{eqnarray}
X^{2}&=&\frac{1}{\alpha^2\rho^2}\left(1-\frac{4M}{\alpha^3\rho^3}\right)
\left[l^2+\frac{k^2}{\alpha^2}+\frac{\kappa^2
n^2}{\alpha^2\left(1-\frac{4M}{\alpha^3\rho^3}\right)}\right] \nonumber \\
&+&\frac{9}{4}-\frac{36M^2}{\alpha^6\rho^6}
+\frac{\xi R}{\alpha^2}\left(1-\frac{4M}{\alpha^3\rho^3}\right)
+\frac{X'}{X}\left[\frac{1}{2}\rho+\frac{2M}{\alpha^3\rho^2}
-\frac{16M^2}{\alpha^6\rho^5}\right] \label{eq:x} \\ \label{eq:xevolve}
&+&\left(\frac{X'}{X}\right)^{2}\left[-\frac{3}{4}\rho^2+
\frac{6M}{\alpha^3\rho}-\frac{12M^2}{\alpha^6\rho^4}\right]
+\frac{X''}{X}\left[\frac{1}{2}\rho^2-\frac{4M}{\alpha^3\rho}+
\frac{8M^2}{\alpha^6\rho^4}\right]. 
\nonumber 
\end{eqnarray}
Which can be obtained by substituting (\ref{eq:ansatz}) into (\ref{eq:kg}).
Substituting (\ref{eq:ansatz}) into
(\ref{eq:wronsk}) gives $C=\frac{1}{2}$ for all mode functions.

\qquad A point splitting algorithm developed by Christensen
~\cite{pointsplit} will be used to renormalise the field. In this
technique, one chooses the points $x$ and $x'$ to be nearby points in the
spacetime before the full coincidence limit is taken. It is convenient to
have the points take on equal values of $\rho,\,\varphi$ and $z$ so that
the coordinate separation is given by $\epsilon=\tau-\tau'$. The
unrenormalised Green function now takes on the form: 
\begin{equation}
G_{E}(x,\tau,\tau')=\frac{T}{8\pi^2}\sum_{n=-\infty}^{\infty}e^{\imath n
\kappa\epsilon}\sum_{l=-\infty}^{\infty}\int_{-\infty}^{\infty}dk\:
\Psi_{1}(\rho)\Psi_{2}(\rho) \label{eq:unreng}.
\end{equation}
It should be noted at this point that there exists a superficial 
ultra-violet divergence over $l$ and $k$ in the above expression. This 
divergence can most easily be eliminated using a similar technique as 
Candelas ~\cite{ch:schw} and Anderson ~\cite{and:mass}. It is noted that 
as long as $\tau\neq\tau'$ any multiple of $\delta(\tau-\tau')$ can be 
added to (\ref{eq:unreng}). Substituting $X$ from (\ref{eq:x}) in the 
large 
$l$ and $k$ limit and subtracting this term from (\ref{eq:unreng}) the 
logarithmic divergences can be eliminated giving the following expression:
\begin{eqnarray}
G_{E}(x,\tau,\tau')=\frac{T}{8\pi^2}\sum_{n=-\infty}^{\infty}e^{\imath n
\kappa\epsilon}\left(\sum_{l=-\infty}^{\infty}\int_{-\infty}^{\infty}dk\:
\Psi_{1}(\rho)\Psi_{2}(\rho) \right. \nonumber \\
\left. -\frac{1}{\sqrt{l^2+\frac{k^2}{\alpha^2}}
\sqrt{\alpha^2\rho^2-\frac{4M}{\alpha\rho}}\alpha\rho}\right). 
\label{eq:finiteg} 
\end{eqnarray}

\subsection{Renormalisation}

\qquad To calculate the renormalised value of \phis\ a point splitting 
technique will be used. The DeWitt generalisation to Schwinger's expansion 
is used as an approximation for the Green function. This term will then 
be subtracted from (\ref{eq:finiteg}) and the $x\rightarrow x'$ limit will 
be taken along the shortest geodesic separating the points. The 
DeWitt-Schwinger counter-term is given by:
\begin{equation}
G(x,x')=\frac{1}{8\pi^2\sigma}+\frac{1}{96\pi^2}R_{\mu\nu}
\frac{\sigma^{;\mu}\sigma^{;\nu}}{\sigma} \label{eq:dscounter}
\end{equation}
where $\sigma$ is the "world function" of Synge ~\cite{syngebook} which 
is equal to half of the square of the geodesic distance between two 
points. The points in this case will be $\tau$ and $\tau'$.

\qquad It can be shown, by geodesic expansion (see appendix A), that the 
world function in the spacetime considered here takes on the form 
\begin{eqnarray}
\sigma&=&\frac{1}{2}\left(\alpha^2\rho^2-\frac{4M}{\alpha\rho}\right)
\epsilon^2-\frac{1}{24}\frac{\left(\alpha^3\rho^3-2M\right)^2
\left(\alpha^3\rho^3-4M\right)}{\alpha^3\rho^5}\epsilon^4+O(\epsilon^6), 
\nonumber \\
\sigma^{;\tau}&=&\epsilon-\frac{1}{6}\frac{\left(\alpha^3\rho^3-2M\right)^2
\left(\alpha^3\rho^3-4M\right)}{\alpha^3\rho^5}\epsilon^3+O(\epsilon^5), 
\label{eq:sigmas} \\
\sigma^{;\rho}&=&\frac{\left(\alpha^3\rho^3+2M\right)
\left(\alpha^3\rho^3-4M\right)}{\alpha^2\rho^3}\epsilon^{2}+O(\epsilon^4). 
\nonumber 
\end{eqnarray}
Using these expressions the DeWitt-Schwinger counter term is equal to
\begin{eqnarray}
G_{counter}&=&
\frac{1}{4\pi^2\left(\alpha^2\rho^2-\frac{4M}{\alpha\rho}\right)\epsilon^2}
+\frac{1}{48\pi^2}\frac{\left(\alpha^3\rho^3+2M\right)^2}
{\alpha\rho^3\left(\alpha^3\rho^3-4M\right)} \nonumber \\ 
&-&\frac{3\alpha^2}{48\pi^2}\label{eq:moreds}
\end{eqnarray}

\qquad The first term in (\ref{eq:moreds}) can be rewritten in a more 
convenient way by using the Plana sum formula ~\cite{plana} as used by 
Anderson. The formula is
\begin{equation}
\sum_{n=n_{0}}^{\infty}f(n)=\frac{1}{2}f\left(n_{0}\right)+
\int_{n_{0}}^{\infty}f(s)\,ds+\imath\int_{0}^{\infty}
\frac{dt \: \left[f\left(n_{0}+\imath t\right)-
f\left(n_{0}- \imath t\right)\right]}{e^{2\pi t}-1}.
\end{equation}
With this, the first expression in the counter term can be written as
\begin{eqnarray}
\frac{1}{4\pi^{2}\left(\alpha^2\rho^2-\frac{4M}{\alpha\rho}\right)\epsilon^2}
= \nonumber \\
-\frac{\kappa}{4\pi^2\left(\alpha^2\rho^2-\frac{4M}{\alpha\rho}\right)}
\sum_{n=1}^{\infty}\cos(n\kappa\epsilon)\,n\,\kappa-
\frac{\kappa^2}{48\pi^2\left(\alpha^2\rho^2-\frac{4M}{\alpha\rho}\right)},
\end{eqnarray}
 which gives, for the entire counter-term in the $\epsilon\rightarrow 0$
limit: 
\begin{eqnarray}
G_{counter}&=&
-\frac{\kappa}{4\pi^2\left(\alpha^2\rho^2-\frac{4M}{\alpha\rho}\right)}
\sum_{n=1}^{\infty}n\,\kappa
-\frac{\kappa^2}{48\pi^2\left(\alpha^2\rho^2-\frac{4M}{\alpha\rho}\right)}
\nonumber \\
&+&\frac{1}{48\pi^2}\frac{\left(\alpha^3\rho^3+2M\right)^2} 
{\alpha\rho^3\left(\alpha^3\rho^3-4M\right)}
-\frac{3\alpha^2}{48\pi^2}. \label{eq:fullcounter}
\end{eqnarray}
It can be shown that the second and third terms in 
(\ref{eq:fullcounter}), which normally diverge at the horizon, will 
cancel each other out on the horizon when $T$ is equal to the black hole 
temperature. That is, when
\begin{equation}
T=\frac{\alpha}{2\pi}\frac{3}{2}\left(4M\right)^{\frac{1}{3}}.
\end{equation}
Therefore, the renormalised expression for \phis\ in the Hartle-Hawking 
vacuum state is
\begin{eqnarray}
\langle\phi^2\rangle&=&
\frac{T}{8\pi^2}\left(\sum_{n=-\infty}^{\infty}
\sum_{l=-\infty}^{\infty}\int_{-\infty}^{\infty}dk\: 
\Psi_{1}(\rho)\Psi_{2}(\rho) \right.  
\left.-\frac{1}{\sqrt{l^2+\frac{k^2}{\alpha^2}}
\sqrt{\alpha^2\rho^2-\frac{4M}{\alpha\rho}}\alpha\rho}\right)
\nonumber \\
&+&\left(\frac{\kappa}{4\pi^2\,f}\sum_{n=1}^{\infty}n\,\kappa\right)
+\frac{\kappa^2}{48\pi^{2}\, f}-\frac{f^{\prime\,2}}{192\pi^2\, f}
+\frac{3\alpha^2}{48\pi^2}. \label{eq:phisq}
\end{eqnarray}
Where, for convenience, the following notation has been used
\begin{eqnarray}
f&\equiv&g_{00}=\left(\alpha^2\rho^2-\frac{4M}{\alpha\rho}\right), 
\nonumber \\ f^{\prime}&=&\frac{\partial f}{\partial \rho}.
\end{eqnarray}

\section{Calculation of \phis}

\qquad In this section the value of \phis\ will be calculated from
(~\ref{eq:phisq}). The solution is found by iteratively solving for the
function $X$ using (\ref{eq:xevolve}) in the mode functions with the lowest
order term defined as: 
\begin{equation}
X_{0}^{2}=\frac{h}{\alpha^2\rho^2} 
\left[l^2+\frac{k^2}{\alpha^2}+\frac{\kappa^2 n^2}{\alpha^2\, h}\right]
+\frac{9}{4}-\frac{36M^2}{\alpha^6\rho^6}+\frac{\xi R}{\alpha^2}h,
\end{equation}
and $h$ given by 
\begin{equation}
h=\left(1-\frac{4M}{\alpha^3\rho^3}\right). 
\end{equation}
With this choice, the lowest order term will be valid at both the horizon 
and as $\rho$ approaches infinity. This can also be verified by comparison 
with (\ref{eq:infbehav}) and (\ref{eq:horizbehav}).

\qquad It is most convenient to do the integration over $k$ first 
followed by the 
sum over $l$ and finally, the sum over $n$. This scheme leads to analytic 
solutions to both the integrals and the $l$ sums. 

\qquad In the solution, the sums and integrals have 
the following form: 
\begin{equation}
\sum_{n}\sum_{l}\int_{0}^{\infty}\frac{dk}{\left(l^2+\frac{k^2}{\alpha^2}+
\frac{\kappa^2 n^2}{\alpha^2 h}+V_{0}(\rho)\right)^{\frac{p}{2}}},
\end{equation}
where $V_{0}(\rho)$ is a function of $\rho$ only.
Such integrals are known and are given by
\begin{equation}
\int_{0}^{\infty}\frac{dk}{\left(l^2+\frac{k^2}{\alpha^2}+
\frac{\kappa^2 n^2}{\alpha^2 h}+V_{0}(\rho)\right)^{\frac{p}{2}}}
= C\frac{\alpha}{\left(l^2+\frac{\kappa^2 n^2}{\alpha^2 
h}+V_{0}(\rho)\right)^ {\frac{p-1}{2}}},
\end{equation}
where $C$ is a fractional constant which depends on the particular value 
of the integer $p$. The resulting sums over $l$ can now be done 
analytically by the standard contour integration:
\begin{equation}
\sum_{-\infty}^{\infty}\frac{1}{\left(l^2+\frac{\kappa^2 n^2}{\alpha^2 
h}+V_{0}(\rho)\right)^ {\frac{p-1}{2}}}=-\sum\left(\hbox{Residues of }\:
\frac{\pi\cot\pi\,l}{\left(l^2+\frac{\kappa^2 n^2}{\alpha^2 
h}+V_{0}(\rho)\right)^ {\frac{p-1}{2}}}\right).
\end{equation}
This is valid as the denominator never becomes singular at integer values 
of $l$.

\qquad In the expansion of \phis\ there appears a term of the form
\begin{equation}
\frac{1}{\rho^2\alpha\sqrt{h}}\ln\left(\frac{|l|}{\sqrt{l^2+\frac{\kappa^2 
n^2} {\alpha^2\,h}+V_{0}(\rho)}}\right) \label{eq:lterm},
\end{equation}
which arises when the lowest order term is combined with the counterterm 
in (\ref{eq:finiteg}) and integrated over $k$.
This is the only term which is singular at $l=0$. The $l=0$ term is 
therefore removed from the sum and redefined as to remove a spurious 
divergence of the form 
$\ln\left(2\pi\sqrt{\frac{\kappa^2\,n^2}{\alpha^2 h}+V_{0}(\rho)}\right)$ 
which is present 
in the same term once the sum from $l=1$ to $\infty$ has been calculated 
(how this term arises is shown in more detail in appendix B.) 
This redefinition can be performed since the counterterm introduced in 
(~\ref{eq:finiteg}) may be any quantity and the particular value shown 
there is exact only in the large $l$ limit.

\qquad The final sum over $n$ can now numerically be shown to 
converge by computing the values of \phis\ for large $n$. It can also be 
shown that the $n=0$ mode makes no contribution to the sum.

\qquad The boundary values of the modes must be evaluated before numerical 
integration of the mode equation can be done.
The value of the mode functions at infinity can 
easily be seen by studying the mode equation in the asymptotic region and 
therefore only the value at the horizon is left to be determined. At the 
horizon, there are many quantities in the expansion of \phis\ which are 
inversely proportional to some power of the metric function $f$. By 
performing an expansion of \phis\ in the quantity $\delta=\rho-\rho_{H}$, 
where $\rho_{H}$ is the horizon value of $\rho$, 
one can show (although the procedure is lengthy) that all terms with
$\delta$ raised to some power in the denominator cancel at the horizon.
In appendix B it is also demonstrated how terms which 
normally make a dominant contribution to \phis\ cancel here. It 
should be noted 
that the horizon value is directly proportional to the value of the 
cosmological 
constant. This is due to the fact that the spacetime is an Einstein 
spacetime with $R_{\mu\nu}=3\alpha^2\,g_{\mu\nu}$.

\qquad \phis\ was computed for the conformally coupled, massless case and 
the result is shown in Fig.1. It can be seen the the maximum value of 
\phis\ occurs near, but not at, the horizon. This behaviour is analogous 
to the extreme Reissner-Nordstr\"{o}m case \cite{and:mass}. This is 
because, as shown in the appendix and earlier, 
most contributions to \phis\ at the horizon vanish. 
However, near (but not on) the horizon terms with a $1/f$ behaviour make a 
large contribution. For large $\rho$ most terms in the field expansion 
vanish and therefore \phis\ approaches a value
which is dominated by the last two terms in (\ref{eq:phisq}).

\begin{figure}[ht]
\includegraphics[bb= 26 39 541 455,width=0.6\textwidth,clip]{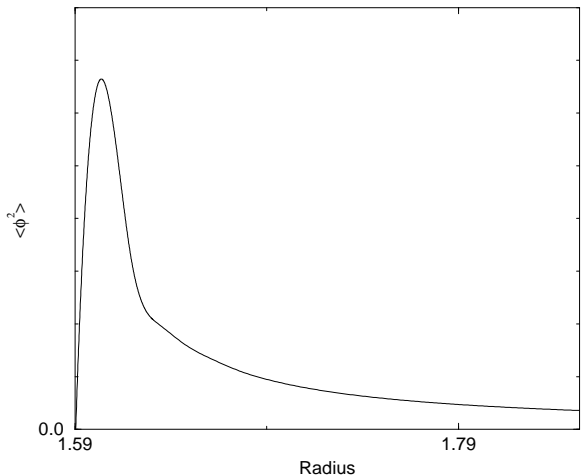}
\caption{\label{fig1} \phis\ for the cylindrical black hole spacetime. 
The value of \phis\ has a small but nonzero value at the horizon and 
attains a maximum away from the horizon.}
\end{figure}

\section{Acknowledgements} 
\qquad The author would like to thank Dr. K. S. Viswanathan for advice and 
help without which this work 
would not have been possible. The author would also like to thank Dr. P. 
Anderson of Wake Forest University for his patience in answering 
questions via e-mail. Dr. N. Kaloper was also very helpful in drawing my 
attention to some references which were originally missed. 

\section{Appendix A}
\qquad In this appendix it will be demonstrated how to calculate the 
world function, $\sigma$, using the method of geodesic expansion. 
Equations here involve quantities which are defined at different 
spacetime points. For a brief review of handling bitensors the reader is 
referred to Christensen ~\cite{pointsplit} and Synge ~\cite{syngebook}.

Let $P_{1}(x)$ and $P_{2}(x')$ be two points in the spacetime close enough 
together such that they are connected by a {\it unique} geodesic. The 
geodesic equation
\begin{equation}
\frac{d^{2}x^{\mu}}{ds^{2}}+\Gamma^{\mu}_{\alpha\beta}
\frac{dx^\alpha}{ds}\frac{dx^\beta}{ds}=0
\end{equation}
yields the power series
\begin{eqnarray}
x^{\mu}&\approx& x^{\mu'}+U^{\mu'}\,ds-\frac{1}{2}\Gamma^{\mu'}_{\alpha' 
\beta'}U^{\alpha}U^{\beta}ds^2+\frac{1}{6}\left(2\Gamma^{\mu'}_{\gamma' 
\delta'}\Gamma^{\gamma'}_{\kappa'\eta'}U^{\kappa}U^{\eta}
U^{\delta}\right) ds^3 \nonumber \\
&-&\frac{1}{6}\Gamma^{\mu'}_{\kappa'\eta',\gamma'}
U^{\kappa}U^{\eta}U^{\gamma}ds^3 + ... \label{eq:geoexp}
\end{eqnarray}
where
\begin{equation}
U^{\mu}=\frac{dx^{\mu}}{ds}
\end{equation}
and
\begin{equation}
U^{\mu}ds=dx^{\mu}\approx\xi^{\mu}=(\tau-\tau')\delta^{\mu}_{0}.
\label{eq:xidef}
\end{equation}
The last term in equation (\ref{eq:geoexp}) is zero since the spacetime 
is static. (\ref{eq:geoexp}) can be inverted and used in the definition 
~\cite{syngebook} \begin{equation}
2\sigma(x,x')=ds^2\,g_{\mu'\nu'}U^{\mu'}U^{\nu'}
\end{equation}
giving
\begin{eqnarray}
2\sigma(x,x')&=&\xi^{\mu}\xi^{\nu} g_{\mu'\nu'}+\frac{1}{2}\xi^{\mu}
\xi^{\epsilon}\xi^{\kappa}\Gamma^{\nu'}_{\epsilon'\kappa'}g_{\mu'\nu'}
-\frac{2}{3}\xi^{\omega}\xi^{\xi}\xi^{\rho}\xi^{\mu}
\Gamma^{\nu'}_{\chi'\rho'}\Gamma^{\chi'}_{\omega'\xi'}g_{\mu'\nu'} 
\nonumber \\ 
&+&\frac{1}{2}\xi^{\alpha}\xi^{\beta}\xi^{\nu}
\Gamma^{\mu'}_{\alpha'\beta'}g_{\mu'\nu'}+\frac{1}{4}
\xi^{\alpha}\xi^{\beta}\xi^{\epsilon}\xi^{\sigma}
\Gamma^{\mu'}_{\alpha'\beta'}\Gamma^{\nu'}_{\epsilon'\sigma'}
g_{\mu'\nu'}.
\end{eqnarray}
Calculating the Christoffel symbols using (\ref{eq:metric}) and noting 
(\ref{eq:xidef}) it can be shown that this expression reduces to the one in
(\ref{eq:sigmas}).
This expression has the same functional form as that of the world 
function for a static, spherically symmetric spacetime ~\cite{and:mass} 
if the coordinate separation there is also chosen as (\ref{eq:xidef}). 
This is due to the fact that the $\xi^\mu$ vectors eliminate any 
dependence of $\sigma(x,x')$ on $g_{22}$ and $g_{33}$.

\section{Appendix B}
\qquad In this appendix the dominant terms of \phis\ at the horizon will be 
calculated. This is useful since the value at the horizon needs to be 
evaluated as a starting point for the calculation of the mode functions. 
This calculation is also useful as it provides insight as to how the $n$ 
counter-term acts to regularise the field.

\qquad At the horizon, the dominant terms in the field expansion are 
given by 
\begin{equation}
\langle\phi^2\rangle=\frac{T}{4\pi^2}\sum_{n=-\infty}^{\infty}
\sum_{l=-\infty}^{\infty}\frac{1}{\rho^2\alpha\sqrt{h}}\ln\left(
\frac{|l|}{\sqrt{l^2+\frac{\kappa^2 n^2}{\alpha^2 h}+V_{0}(\rho)}}\right)
+\frac{\kappa}{4\pi^2 f}\sum_{n=0}^{\infty}n\kappa. \label{eq:horizphi} 
\end{equation} 
For the
moment, we choose to ignore the $l=0$ term and concentrate on the first
expression in (\ref{eq:horizphi}) which can trivially be re-written as
\begin{eqnarray} 
&=&\frac{T}{2\pi^2}\sum_{n=-\infty}^{\infty}
\sum_{l=1}^{\infty}\frac{1}{\rho^2\alpha\sqrt{h}}\ln\left(
\frac{1}{\sqrt{1+\frac{\frac{\kappa^2 n^2}{\alpha^2 
h}+V_{0}(\rho)}{l^2}}}\right) \nonumber \\ 
&=&-\frac{T}{4\pi^2}\sum_{n=-\infty}^{\infty}
\sum_{l=1}^{\infty}\frac{1}{\rho^2\alpha\sqrt{h}}\ln\left(
1+\frac{\frac{\kappa^2 n^2}{\alpha^2 h}+V_{0}(\rho)}{l^2}\right) 
\nonumber \\ 
&=&-\frac{T}{4\pi^2}\sum_{n=-\infty}^{\infty}\frac{1}
{\rho^2\alpha\sqrt{h}}\ln\left(\prod_{l=1}^{\infty}\left(1+\frac{
\frac{\kappa^2 n^2}{\alpha^2 h}+V_{0}(\rho)}{l^2}\right)\right). 
\label{eq:prod} \end{eqnarray} 
The product in the above expression is well known yielding the following for
(\ref{eq:prod}) 
\begin{eqnarray}
&-&\frac{T}{4\pi^2}\sum_{n=-\infty}^{\infty}\frac{1}
{\rho^2\alpha\sqrt{h}}\left[\ln\left(2\sinh\left(\sqrt{\frac{\kappa^2 n^2}
{\alpha^2 h}+V_{0}(\rho)}\,\pi\right)\right) \right. \nonumber \\
&-&\left.\ln\left(2\pi\sqrt{\frac{\kappa^2
n^2}{\alpha^2 h}+V_{0}(\rho)}\right) \right]. \label{eq:nleft}
\end{eqnarray} 
For very large $n$ or (as in the case here) very small $h$ this becomes 
\begin{equation} 
-\frac{T}{4\pi^2}\sum_{n=-\infty}^{\infty}\frac{1}
{\rho^2\alpha\sqrt{h}}\left[\frac{\kappa |n| \pi}
{\alpha\sqrt{h}}-\ln\left(\frac{2 \kappa |n|\pi}{\alpha\sqrt{h}}\right) 
\right]. \label{eq:lzero} 
\end{equation}
If we define the $l=0$ term to cancel out the second term in 
(\ref{eq:nleft}) as described in the text, the resultant expression gives
\begin{equation}
-\frac{2\pi T}{4\pi^2\,f}\sum_{n=0}^{\infty}n\kappa
\end{equation}
which is cancelled by the $n$ sum in the counter-term. This leaves a 
small constant contribution to \phis\ at the horizon.

\newpage
\bibliographystyle{unsrt}

\end{document}